\documentclass{aa}
\usepackage{graphics}
\begin{document}

\thesaurus{03.13.1; 08.01.1; 10.19.1}

\title{A possible solution of the G-dwarf problem in the frame-work of
closed models with a time-dependent IMF}

%ago   \subtitle{I. Overviewing the $\kappa$-mechanism}

   \author{Agostino Martinelli
          \inst{1}
          \and
          Francesca Matteucci
          \inst{2}
          }
\offprints{A. Martinelli}

\institute{
Dipartimento di Astronomia, Universit\`a di Trieste, Via G.B. Tiepolo, 11,
34131 Trieste, Italy \and
SISSA/ISAS, Via Beirut 2-4, 34014 Trieste, Italy}

   \date{Received ; accepted }

\titlerunning{The G-dwarf problem}
\authorrunning{Martinelli \& Matteucci}

   \maketitle

\newcommand{\be}{\begin{equation}}
\newcommand{\ee}{\end{equation}}

\begin{abstract}

In this paper we present a method to solve the G-dwarf problem in the
frame-work of analytical models (based on the instantaneous recycling 
approximation, IRA). We consider a
one-zone closed model without inflows or outflows. We suppose a
time-dependent Initial Mass Function (IMF)
and we find an integral-differential equation which 
must be satisfied
in order to honour the G-dwarf metallicity distribution as a 
function of the oxygen
abundance.
IMFs with one and two slopes are given and discussed
also in the framework of a numerical chemical evolution 
model without IRA. We conclude that it is difficult
to reproduce other observational constraints besides
the
G-dwarf distribution (such as $[\frac{O}{Fe}]$ vs 
$[\frac{Fe}{H}]$),
and that an IMF with two slopes, with time-dependent shape
at the low mass end, would be required.
However, even in this case the predicted oxygen gradient along the disk is flat
and radial flows would be required to reproduce the observed gradient.

\keywords{solar neighbourhood - stars: abundances - methods: analytical}

\end{abstract}

\section{Introduction}

It is well known that the metallicity
distribution of G-dwarfs in the solar neighbourhood shows a deficit of 
metal-poor stars relative to the predictions of the Simple Model of
chemical evolution. This is the so-called G-dwarf problem, originally
noted by van den Bergh (1962) and Schmidt (1963).

Many solutions have been proposed, some in the frame-work of 
analytical models, based on the IRA. The problem
is still present if we adopt the oxygen abundances in place of metallicity.

An important result in the frame-work of IRA models with gas flows
was pointed out by Edmunds (1990). He found that the G-dwarf problem
cannot be solved by any outflow but it is possible to solve it
with particular forms of inflow. Lynden Bell (1975) 'best accretion model'
and Clayton's models (Clayton 1988) are models of just this kind.

Of course it is also possible to reduce the number of metal-poor stars
with respect to that predicted by the Simple Model
by assuming metal dependent stellar
yields (Maeder 1992) (more metals are produced
at lower metallicity with a consequent Prompt
Initial Enrichment (P.I.E.) (Truran and Cameron 1971)). However,
it has been shown by several papers
(Giovagnoli and Tosi, 1995; Carigi 1996), that in this way shallow
gradients along the Galactic disk
are produced, which do not agree with the most
recent observational estimates
(see Matteucci and Chiappini 1999 for a review).
Actually, this is an unavoidable problem
still present in the models discussed in this paper where
we consider time-dependent IMFs.

It is still possible to obtain a Prompt
Initial Enrichment by assuming
an IMF variable with time and in particular very flat at early times,
in order to favour the formation of massive stars.

In this paper we exploit the previous idea to investigate the time behaviour of the
IMF. We assume the same hypothesis of the Simple Model (Tinsley 1980)
with the exception
of adopting a time-dependent IMF. In such a way it is possible to
find an equation (Section 2.1)
which must be satisfied in order to reproduce the observed
distribution
of G-dwarfs.
We then use this equation to investigate the time behaviour
of IMFs with one and two slopes.
In Section 3 we use a numerical model without IRA
to test the IMFs on other observational constraints.

%
%________________________________________________________________
\section{Recovering the history of the IMF}

In the following we need an analytical expression to
approximate the data concerning the
metallicity distribution of G-dwarfs in the solar neighbourhood. The data are
plotted in Fig. 1 where the differential distribution of oxygen 
abundances for the
solar cylinder from Rocha-Pinto and Maciel (1996) is given (triangles).

We have made use of the oxygen abundance rather than the iron abundance, 
in order
to be consistent with chemical evolution models which use the instantaneous
recycling approximation.  
Indeed, IRA is a good approximation for oxygen which is produced on short
timescales (few million years) by supernovae (SNe) II  as opposed to iron
which is produced on long timescales (up to a Hubble time) by SNe Ia
(Matteucci and Greggio 1986).
In order to compare theory and data we adopt 
a simple relation between O and Fe  reproducing the observed
behaviour of these elements in the solar neighbourhood,
the same  
as 
given by Pagel (1989):

%eq1
\be
log(\Phi)=\left[\frac{O}{H}\right]=0.5\left[\frac{Fe}{H}\right]
\ee

\begin{figure}
\resizebox{\hsize}{!}{\includegraphics{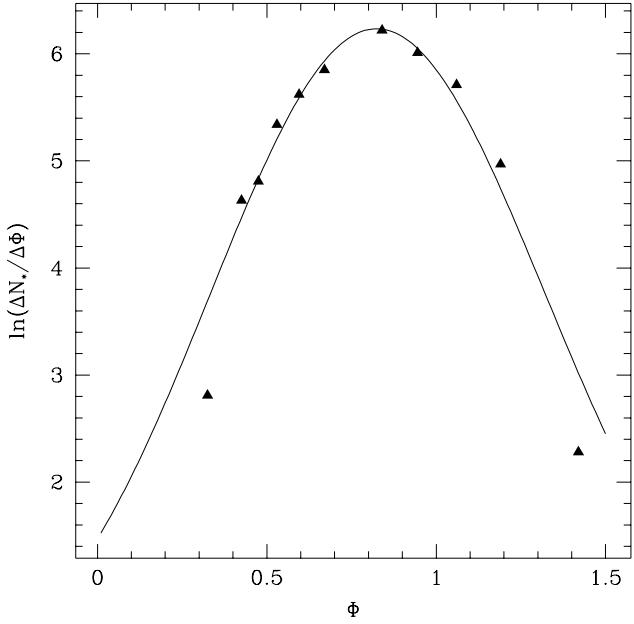}}
\caption{Metallicity distribution of 287 G-dwarf stars, from Rocha-Pinto and Maciel
1996 (triangles) and the function $f(\Phi)$ (eq. 2) (continuous line).
$\Phi$ is the oxygen abundance in units of the solar abundance.
$\Delta N_*$ is the number of stars with oxygen abundance between $\Phi$ and 
$\Phi+\Delta \Phi$}
\end{figure}

\noindent
The continuous line in Fig. 1 is the best function approximating data points
according to the theory discussed in the appendix.

%eq2
\be
f(\Phi)=ln\left(\frac{\Delta N_*}{\Delta \Phi}\right)=
w_1 e^{-\frac{(\Phi-\Phi_1)^2}{2\sigma^2}}+
w_2 e^{-\frac{(\Phi-\Phi_2)^2}{2\sigma^2}}
\ee

\noindent
with

%eq3
\be
\sigma=0.454;~
\Phi_1=0.690;~
\Phi_2=1.06;~
w1=4.12;~
w2=2.61;
\ee

\noindent
We have used the sophisticated method in the appendix
to approximate the data
because of the strong dependence of our results on the chosen
function $f(\Phi)$ and in particular on the value at $\Phi=0$,
$f(0)$.
Indeed, the point 
$\Phi=0$ is outside of the range of the available data and the
metallicity distribution extrapolated to $\Phi=0$ depends strongly
on the adopted method.
Since the aim of this paper is to recover the history of
the IMF from the data in Fig. 1, we need a method
that does not introduce any {\it a priori} assumption
on the specific functional form of the relation
$ln(\Delta N_*/ \Delta \Phi)$.
In other words our approach must be independent of any physical
hypothesis on the star formation law
at the basis of the observed data,
and this is just the essence of the method
discussed in the appendix.

\subsection{Basic Equations}

The basic assumptions of our models are the same of the Simple Model (Tinsley 1980)
with the only exception of adopting a time-dependent IMF. Therefore we consider a
model with the following properties:

\begin{enumerate}

\item The system is one-zone and closed, namely there are no inflows or outflows

\item The initial gas is primordial (no metals)

\item The gas is well mixed at any time

\item The instantaneous recycling approximation holds (namely the lifetimes of stars
above $1M_{\odot}$ are negligible whereas those of stars below $1M_{\odot}$ are
larger than the age of the Galaxy).

\item The IMF is time-dependent, i.e. $\varphi=\varphi(m,t)$,
with the following normalization at any time

%eq4
\be
\int_0^{\infty}\varphi(m,t)mdm=1
\ee

\end{enumerate}

\noindent
Under the previous assumptions the oxygen abundance 
$\Phi$ (defined by eq.(1)) in the
interstellar medium is governed by

%eq5
\be
d\Phi=\frac{p}{gZ_{0}}\alpha ds
\ee

\noindent
where $Z_{0}=H\frac{O_{\odot}}{H_{\odot}}\simeq 9.54 \cdot 10^{-3}$,
$g$ is the gas mass, $p \alpha ds=p' ds$ is the mass of oxygen 
produced and
almost immediately returned into the interstellar 
medium when $ds$ of interstellar
material goes into stars. A fraction $\alpha$ of the mass gone into stars is
not returned to the interstellar medium, but remains in long-lived stars or
stellar remnants. It has also been assumed that $\Phi Z_{0}<<1$. We have:

%eq6
\be
\alpha=\int_0^{M_{\odot}}\varphi(m,t)mdm+\int_{M_{\odot}}^{\infty}\varphi(m,t)
m_{rem}dm
\ee

\noindent
and

%eq7
\be
p'_{oxy}=\int_{M_{\odot}}^{\infty}\varphi(m,t)mp_{o}dm
\ee

\noindent
where $p_o$ is the fraction (by mass) of newly produced and ejected oxygen by a
star of mass $m$. We used for $p_o$ the expression given by Woosley and Weaver
(1995) whereas for $m_{rem}$ the expression given by Tinsley (1980).

In our model the gas and stellar masses are related by

%eq8
\be
dg=-\alpha ds
\ee

\noindent
Now we want to relate 
the previous quantities to the observed metallicity distribution
of G-dwarfs ($f(\Phi)$ in eq. (2)) in order to find an
equation for $\varphi(m,t)$.
We have:

%eq9
\be
f(\Phi)=ln\left(\alpha_c \frac{ds}{d\Phi}\right)
\ee

\noindent
where

%eq10
\be
\alpha_c=\int_{0.8M_{\odot}}^{1.1M_{\odot}}\varphi(m,t)mdm
\ee

\noindent
because the stars in our sample are in the range $0.8-1.1M_{\odot}$.

Equations (5) and (9) give:

%eq11
\be
f(\Phi)=ln\alpha_c-ln\alpha+lng-lnp+lnZ_{0}
\ee

\noindent
On the other hand we can derive $lng$ from (5) and (8) in
the following way:

%eq12
\be
lng=lng_0-Z_{0}\int_0^{\Phi}\frac{d\Phi}{p}
\ee

\noindent
where $g_0$ is the initial gas mass. Hence from (11) we obtain:

%eq13
\be
f(\Phi)=ln\alpha_c-ln\alpha-
Z_{0}\int_0^{\Phi}\frac{d\Phi}{p}+lng_0-lnp+lnZ_{0}
\ee

\noindent
If we evaluate the previous equation at $\Phi=0$ (that is $t=0$) we have:

%eq14
\be
f(0)=ln\alpha_{c0}+lng_0-lnp'_0+lnZ_{0}
\ee

\noindent
where $p'_0$ and $\alpha_{c0}$ are the quantities in (6) and (10), 
respectively,
with $\varphi(m,t)=\varphi(m,0)$.

By differentiating eq. (13) with respect to $\Phi$ we obtain finally

%eq15
\be
\frac{d}{d\Phi}\left[ln\alpha_c-lnp'\right]-\frac{\alpha}{p'}Z_{0}=
\frac{df(\Phi)}{d\Phi}
\ee

\noindent
This is an integro-differential equation for the function $\varphi(m,\Phi)$ with
the initial condition given by (14).
The previous problem has, of course, infinite
solutions. Therefore to proceed further
we have to make some assumption on the behaviour of
the $\varphi(m,\Phi)$. In the next sections we shall investigate
IMF with one slope (Sect. 2.2) and with two slopes (Sect. 2.3).

%It is interesting to note that eq. (15)
%becomes a quite simple integral equation if it is possible to assume the relation
%$\alpha_c\simeq\alpha$. In this case equation (15) is a I order differential 
%equation for $p$ with the
%solution $p(\Phi)=[g_0-\int_0^{\Phi}e^{f(\Phi')d\Phi'}]e^{-f(\Phi)}Z_{0}$. This
%is an integral equation for $\varphi$ much simpler than (15).

\subsection{Single Power-law IMF}

Let consider a single power-law IMF, namely

%eq16
\be
\varphi(m,\Phi)= C m^{-[1+x(\Phi)]}
\ee

\noindent
where $M_L$ and $M_U$ are respectively the smallest and the largest stellar mass
(that
we assume do not depend on the time) and the normalization
is performed in the above mass range, i.e.
$C=\frac{1-x(\Phi)}
{M_U^{1-x(\Phi)}-M_L^{1-x(\Phi)}}$

Substituting eq. (16) in (15) we find the following equation for $x(\Phi)$:

%eq17
\be
\frac{dx}{d\Phi}=\frac{F_2(x)+F_3(\Phi)}{F_1(x)}
\ee

\noindent
which
is a nonlinear I order differential equation with initial condition given by
substituting (16) in (14). The functions in (17) are:

%eq18
\be
F_1(x)=\frac{d}{dx}[ln\alpha_c -lnp']
\ee

%eq19
\be
F_2(x)=\frac{\alpha}{p'}Z_{0}
\ee

%eq20
\be
F_3(\Phi)=
\frac{df}{d\Phi}=\\
-\left[
w_1\frac{\Phi-\Phi_1}{\sigma^2} e^{-\frac{(\Phi-\Phi_1)^2}{2\sigma^2}}+
w_2\frac{\Phi-\Phi_2}{\sigma^2} e^{-\frac{(\Phi-\Phi_2)^2}{2\sigma^2}}
\right]
\ee

\noindent
Since $F_2(x)>0$, eq. (17) tells us that a constant IMF (i.e. with a
slope $x=x_0$ at any time) corresponds to a straight line for $f(\Phi)$ with a
negative slope
(Indeed, in this case eq. (17) becomes 
$F_3(\Phi)=\frac{df}{d\Phi}=-F_2(x_0)$).
This is the right behaviour since our model becomes the Simple
Model when $x=x_0$.

The solutions $x(\Phi)$ are plotted in Fig. 2 for two different values of $M_U$ and
$M_L$.
We found very flat IMFs for low oxygen abundances (i.e. at initial times). At the
solar metallicity ($\Phi=1$) the slope is always steeper than a Salpeter (1955)
($x=1.35$).
Decreasing $M_U$ decreases, of course, $x$, i.e. the IMF becomes flatter.
The effect of a change in the $M_L$ value is negligible (especially at low 
$\Phi$).

\begin{figure}
\resizebox{\hsize}{!}{\includegraphics{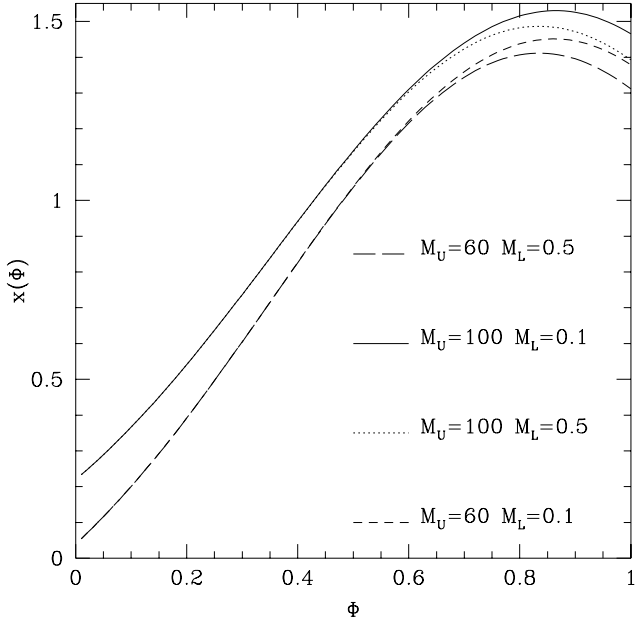}}
\caption{IMF slope ($x(\Phi)$) as a function of oxygen abundance 
($\Phi$) for
two choices of the parameters $M_U$ and $M_L$. The values of $M_U$ and 
$M_L$ are given
in solar mass.}
\end{figure}

\subsection{IMF with two slopes}

Let now consider the following IMF:

\[\varphi(m,\Phi) =  C \left\{ \begin{array}{ll}
m^{-[1+x_1]} & \mbox{if $m \leq M$} \\
M^{x_2-x_1}m^{-[1+x_2]} & \mbox{if $m>M$}
\end{array}
\right. \]

The above IMF depends on the three parameters $x_1, x_2$ and $M$.
We shall investigate the three following cases:

\begin{enumerate}

\item $M=M(\Phi)$, $x_2=1.35$ for
values of $x_1$ ranging from $-1$
to $0.2$ (this IMF is
similar to the one proposed by Larson (1998));

\item $x_1=x_1(\Phi)$, $x_2=1.35$ for values
of $M$ ranging from $2$ to $10 M_{\odot}$;

\item $x_2=x_2(\Phi)$, for
values of $x_1$ ranging from $-1$
to $1$.

\end{enumerate}

In all the previous cases equation (15) gives us
a nonlinear I order differential equation for the
$\Phi$-dependent parameter, with the initial conditions given
by the equation (14).
We consider always $M_U=100M_{\odot}$ and
$M_L=0.1M_{\odot}$ since the dependence on these
parameters is negligible.

\begin{figure}
\resizebox{\hsize}{!}{\includegraphics{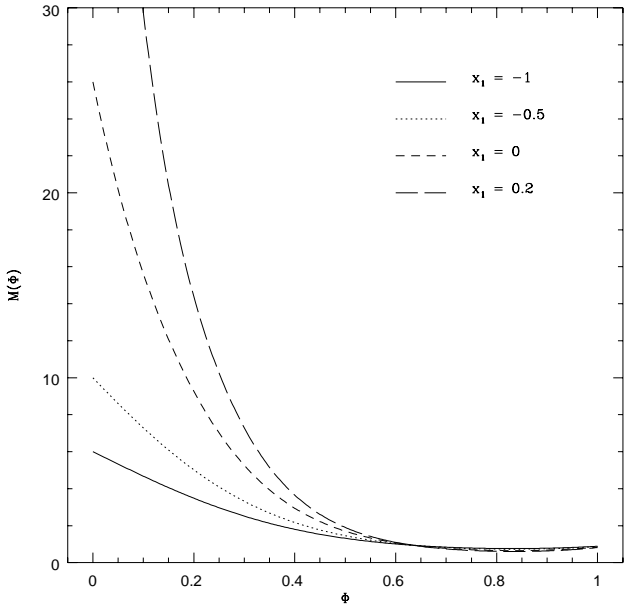}}
\caption{$M(\Phi)$ in solar masses as a function of
$\Phi$ for different $x_1$. The slope $x_2$ is always
$1.35$.}
\end{figure}

The function  $M(\Phi)$ in the first case considered
is shown in Fig. 3. At the initial time ($\Phi=0$) the mass $M$
increases by increasing the slope at low mass end.
In particular when $x_1=0.2$ $M(0)\simeq 100M_{\odot}=M_U$
and therefore there is no solution for $x_1$ greater
than $0.2$.

\begin{figure}
\resizebox{\hsize}{!}{\includegraphics{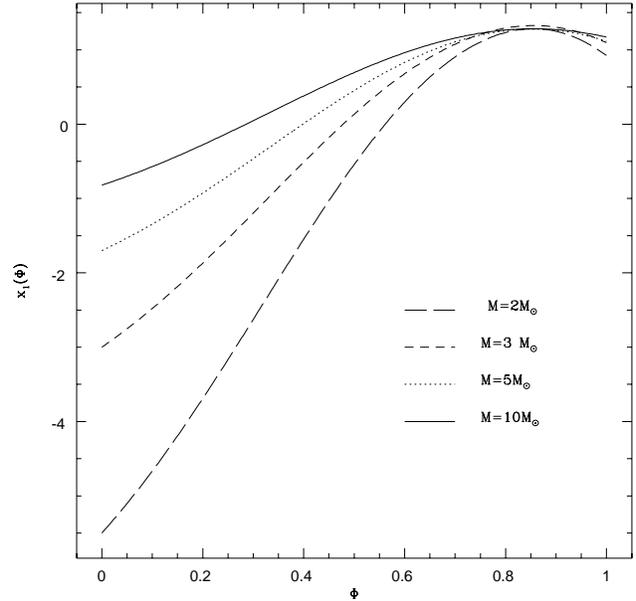}}
\caption{$x_1(\Phi)$ as a function of
$\Phi$ for different $M$. $x_2=1.35$.}
\end{figure}

Fig. 4 shows the slope at the low mass end ($x_1(\Phi)$) for values
of $M$ in the range $2-10 M_{\odot}$.
The initial values of $x_1$ are always negative and moreover they increase
by increasing $M$, as expected.

\begin{figure}
\resizebox{\hsize}{!}{\includegraphics{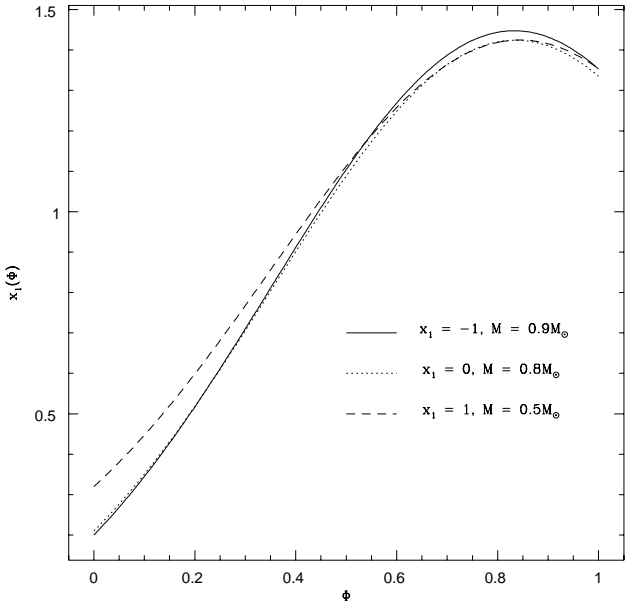}}
\caption{$x_2(\Phi)$ as a function of
$\Phi$ for different $x_1$.
$M$ is chosen in order to have a final slope $x_2 \simeq 1.35$}
\end{figure}

Finally Fig. 5 shows the results for the third case considered.
Here the mass $M$ is chosen in order to reproduce a final
slope $x_2=1.35$.

\section{Application of the derived IMFs to a numerical model}

Both the proposed IMFs can reproduce the observed G-dwarf distribution and
therefore we want
to test the validity of the inferred IMF
by using numerical models of galactic chemical
evolution  . This can be done
by studying the effect of the above IMFs on the chemical evolution of the
solar neighbourhood.
The model used is that of Matteucci and
Francois (1989) where a detailed description can be found. The main
difference with that model is that we assume here a very rapid formation of
either the halo and the disk thus simulating a closed model.
This is required by the fact that we derived the IMF under the assumption
of a closed model. In the original model of Matteucci and Francois (1989)
the timescale for the formation of the solar neighbourhood was about
$3-4$ Gyr and it was chosen in order to best fit the G-dwarf metallicity
distribution of Pagel and Patchett (1975), under the assumption of a
constant IMF. Recently, Chiappini et al. (1997) presented a more realistic
model for the Galaxy where the evolution of the halo-thick disk and the thin
disk are decoupled in the sense that the halo-thick disk is formed on a short
time scale of the order of $1-2$ Gyr, whereas the thin disk is assumed
to form very slowly.

\begin{figure}
\resizebox{\hsize}{!}{\includegraphics{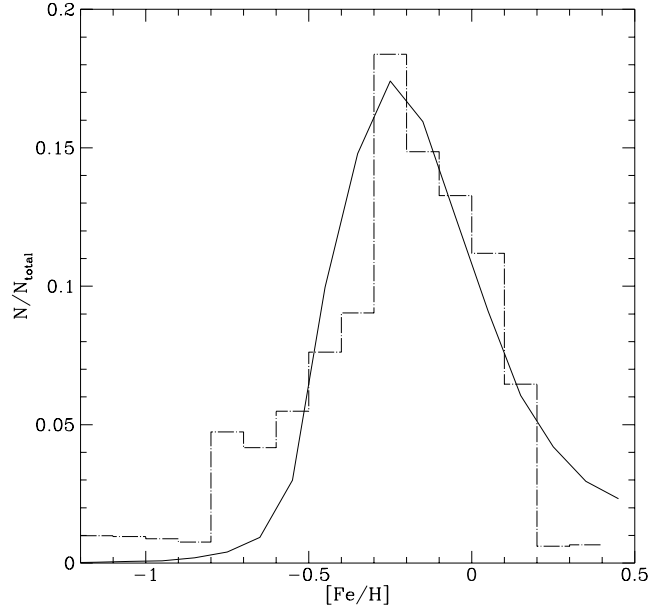}}
\caption{G-dwarf metallicity distribution as predicted by the
numerical model (solid line) with the single
power law IMF discussed in Sect. 2.1, and from experimental data
(Rocha-Pinto and Maciel
1996)}
\end{figure}

\begin{figure}
\resizebox{\hsize}{!}{\includegraphics{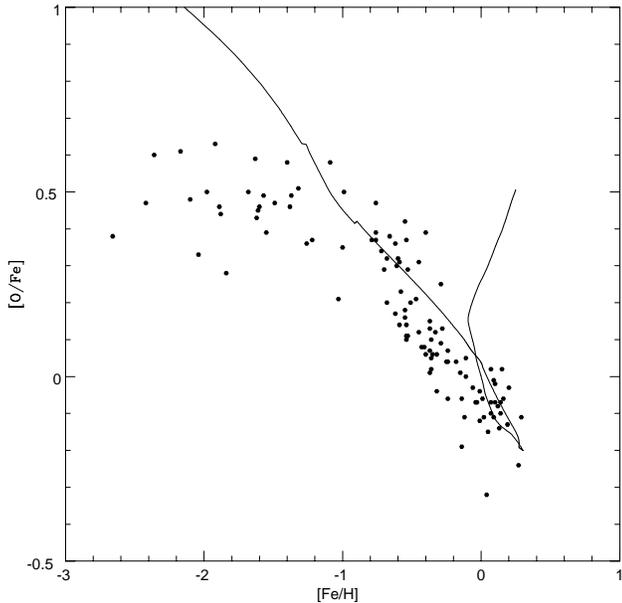}}
\caption{$[O/Fe]$ vs $[Fe/H]$ as predicted by the numerical model (solid line)
with the single
power law IMF (Sect. 2.1),
and as observed}
\end{figure}

In particular, in order to best fit the G-dwarf distribution of
Rocha-Pinto and Maciel (1996), Chiappini et al. (1997) found that, with
an IMF constant in time, a time scale of about $8$ Gyrs is required for
the formation of the solar neighbourhood. It is interesting to check
if the derived IMFs are able to reproduce other observational constraints besides
the G-dwarf distribution, for example the $[O/Fe]$ vs $[Fe/H]$ trend which is
normally very well reproduced by models with constant IMF and infall
and taking into
account detailed nucleosynthesis from type II and Ia SNe. In this framework, in
fact, the plateau shown by the data for $[Fe/H]<-1.0$ is interpreted as due
to the pollution by type II SNe which produce an almost constant $[O/Fe]$
ratio. The subsequent decrease of the $[O/Fe]$ ratio for $[Fe/H]\geq-1.0$ is
then due to the occurrence of type Ia SNe exploding with a temporal delay
relative to SNe II. In the Matteucci and Francois (1989) and
Chiappini et al. (1997) model the SNe Ia are
supposed to originate from white dwarfs in binary systems following the
formalism of Matteucci and Greggio (1986). In Fig. 6 we show the G-dwarf
metallicity distribution as predicted by the numerical model and, as expected,
the agreement is quite good. However, in Fig. 7 we show the predicted
$[O/Fe]$ versus $[Fe/H]$ relation and the agreement with the observations is
poor, especially in the domain of metal poor stars where the new IMF predicts
an increasing amount of massive stars. On the other hand, the slope for
disk stars is well reproduced. Another problem with this IMF is the high
O and 
Fe content reached by the model at the sun birth and at the
present time.  These high abundances are not evident in Fig. 7 since the
abundances are normalized to the predicted solar abundances. 
Finally, another problem is that
Fig. 7 shows that [Fe/H] starts decreasing more than oxygen after having
reached the solar abundance.
This is due to the large dilution from
dying low mass stars and 
to the fact that the Fe abundance
decreases more rapidly than that
of oxygen, due to the fact that
O is continuously produced,
although at a low level, by SNe II on very short timescales.
Iron, on the other hand, comes mostly from type Ia SNe born at early times
when the IMF was top-heavy favoring massive stars relative to the type Ia 
SN progenitors.
The predicted solar
abundances, namely the abundances at 4.5 Gyrs ago, are $X_{O}=2.6 \cdot 10^{-2}$
and $X_{Fe}=6.3 \cdot 10^{-3}$ to be compared with the same abundances from 
Anders and Grevesse (1989) : $X_{O}= 9.59 \cdot10^{-3}$ and $X_{Fe}=1.17 \cdot
10^{-3}$ (mass fractions).

\vskip .2cm

\begin{figure}
\resizebox{\hsize}{!}{\includegraphics{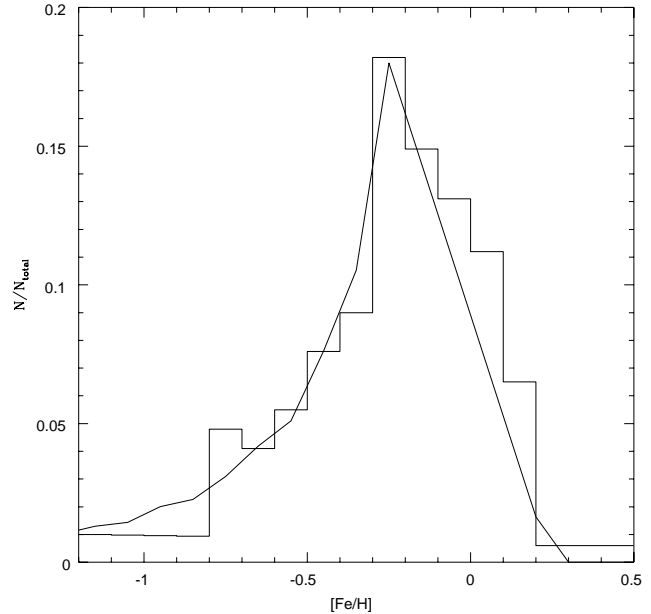}}
\caption{G-dwarf metallicity distribution as predicted by the
numerical model (solid line)
and from experimental data
(Rocha-Pinto and Maciel
1996). The IMF adopted is with two slopes
(discussed in Sect.2.2) with $x_1=x_1(\Phi)$
$x_2=1.35$ and $M=5M_{\odot}$}
\end{figure}

\begin{figure}
\resizebox{\hsize}{!}{\includegraphics{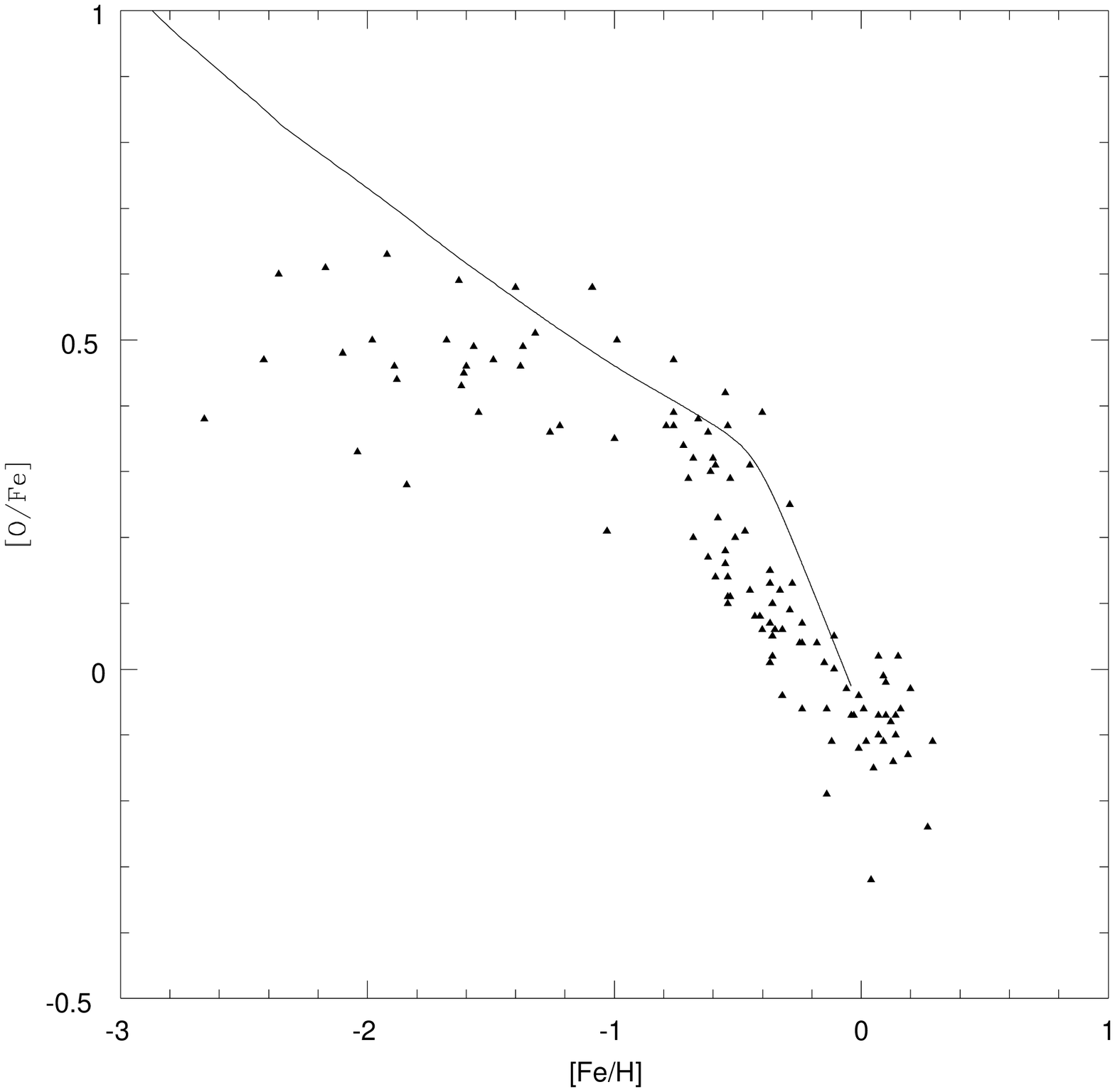}}
\caption{$[O/Fe]$ vs $[Fe/H]$ as predicted by the numerical model (solid line)
and as observed. The adopted IMF is the same as in Fig. 8.}
\end{figure}

\noindent
Concerning the IMF with two slopes (discussed in Sect. 2.3),
the G-dwarf metallicity distribution as predicted by the numerical model
is again in good agreement with the observations, in all
the cases
considered. On the other hand, the predicted
$[O/Fe]$ versus $[Fe/H]$ is still in poor agreement
with the observations 
with the exception of the second case ($x_1=x_1(\Phi)$)
for which we give
the G-dwarf metallicity distribution and the
$[O/Fe]$ versus $[Fe/H]$ in Figs 8 and 9, respectively,
for $M=5M_{\odot}$ (we found results very similar 
to the ones shown in the Figs 8 and 9 for
values of $M$ in the range $2-10 M_{\odot}$).
This time-dependent IMF
fits the $[O/Fe]$ vs $[Fe/H]$ relation
much better than the other time dependent
IMFs, since the variation in the number of type
II SNe at early times is less than in the 
other cases.
However, it is clear that the best agreement with the data
(especially for $[Fe/H]<-1.0$) is achieved with
a constant IMF.

We also investigated the case of an IMF with 
only the lower mass limit dependent on $\Phi$ (($M_L=M_L(\Phi)$). 
However, in this case,
the equations (14) and (15) 
give solutions which are not physical.
In fact, in order to lower the number
of stars in the range of mass
$0.8-1.1 M_{\odot}$ at early times,
to solve the G-dwarf problem, our equations
fix the value of $M_L(\Phi)$ (for $\Phi \simeq 0$)
very close to $1.1 M_{\odot}$.
This result does not have any physical meaning
because it is strongly dependent on the particular mass range
of interest. This is the reason why
the inferred solution, when used in a numerical model
of chemical evolution, it is not able to reproduce the
observational G-dwarf distribution.

The predicted solar
abundances related to the IMF adopted in Figs 8 and 9
are still larger than
the observed ones
($X_{O}=3.1 \cdot 10^{-2}$
and $X_{Fe}=8.6 \cdot 10^{-3}$).
However, these absolute values are strongly model dependent and
in particular it is possible to lower these abundances by increasing the
infall.

\section{Conclusions}

We proposed a method
to solve the G-dwarf problem
in a closed box model with a time-dependent IMF,
based on the IRA. The method
gives us an equation (eq. 15) which has infinite solutions.
Therefore, in order
to use this equation, we had to make some assumptions on the
behaviour of the IMF. 
In particular, we considered both a single power-law IMF
and an IMF with two slopes.

We tested the validity of the inferred IMF
by using numerical models of galactic chemical
evolution, namely by relaxing the 
IRA, and we find the following results:

\begin{enumerate}

\item all the IMFs investigated can
reproduce the observed G-dwarf distribution;

\item a single power-law IMF fails in reproducing
the behaviour of abundances (in particular
the $[O/Fe]$ vs $[Fe/H]$ relation);

\item in order to reproduce the behaviour of abundances
besides the G-dwarf problem,
an IMF with a time dependence at the low
mass end is required.
However, the fit produced by
such an IMF of the 
$[O/Fe]$ vs $[Fe/H]$, is not as good
as that produced by a constant IMF.
Moreover,
this IMF, like most of the variable IMF proposed insofar, fails 
in reproducing the oxygen gradient along the Galactic disk, unless other 
assumptions such as increasing star formation efficiency with galactocentric distance and/or radial flows are introduced.
We computed the expected gradients of oxygen along the disk by adopting this IMF and the model of Chiappini et al. (1997).
We found that the O gradient disappears. A variable efficiency of star formation could recover the gradient but at expenses of the gas distribution
along the disk. We did not try to include radial flows but the conclusion that a variable IMF of this kind worsen the agreement with the disk propeties relative to a constant IMF seems unavoidable.

\item an IMF similar to the one
proposed recently by Larson (1998), fails in reproducing the 
$[O/Fe]$ vs $[Fe/H]$ relation because of the strong
time dependence of
the number of type
II SNe at early times.

\item In summary, from the analysis of models with variable IMF done
in this paper,
we are tempted to conclude that infall models with a constant IMF are
the best solution of the G-dwarf problem, since variable IMFs could 
in principle solve it but they are not able to reproduce the properties 
of the galactic disk. 

\end{enumerate}

%ago \begin{acknowledgements}
%ago       Part of this work was supported by the German
%ago       \emph{Deut\-sche For\-schungs\-ge\-mein\-schaft, DFG\/} project
%ago       number Ts~17/2--1.
%ago \end{acknowledgements}

\section{Appendix}

We give a method to approximate observational data in Fig. 1, based
on the regularization theory of Tikonhov (1963).

The method is general and it applies whenever one wants to approximate
some data by an analytical function, and nothing is known
about the physics at the basis of these data.

Accordingly to Tikonhov's regularization theory (1963) the function
$f(\Phi)$ which 
 approximates the data in Fig. 1 is determined by minimizing
a cost functional $E[f]$, so-called because it maps functions (in
some suitable function space) to the real line. $E[f]$ is the sum of
two terms

%eq21
\be
E[f]=E_s[f]+E_c[f]
\ee

\noindent
where $E_s[f]$ is the standard error term and $E_c[f]$ the regularizing term.
The first-one measures the standard error (distance) between the desired
response $f_i$ and the actual response $f(\Phi_i)$,

%eq22
\be
E_s[f]=\frac{1}{2}\sum_{i=1}^{N}(f_i-f(\Phi_i))^2
\ee

\noindent
where $N$ is the total number of available data (in our case $N=12$).
The second-one depends on the geometric properties of the approximating
function $f(\Phi)$,

%eq23
\be
E_c[f]=\frac{1}{2}\|Pf\|^2
\ee

\noindent
where $P$ is a differential operator.
As suggested by Poggio and Girosi (1990)
the best choice for $P$ consists in a
differential
operator invariant under both rotations and translations.
This is defined by

%A special class of differential
%operators which are invariant under both rotations and translations
%is defined by (Poggio and Girosi 1990):

%eq24
\be
\|Pf\|^2=\sum_{k=0}^{\infty}a_k\|D^k f\|^2
\ee

\noindent
where $a_k=\frac{\sigma^{2k}}{k!2^k}$ with $\sigma$ a constant associated with the
data point $\Phi_i$ and

%eq25
\be
\|D^k f\|^2=\int_{-\infty}^{\infty}\left(\frac{\partial^k f}
{\partial \Phi^k}\right)^2 d\Phi
\ee

\noindent
The function which solves the variational problem given by eq. (21) (when the
regularizing term is specified by eq. (24)) is:

%eq26
\be
f(\Phi)=\sum_{j=1}^{M}w_j e^{-\frac{(\Phi-\Phi_j)^2}{2\sigma^2}}
\ee

\noindent
which consists of a linear superposition of multivariate Gaussian
basis functions with centers $\Phi_j$.
The above theory establishes only the form of the function $f(\Phi)$, but
does not solve completely our problem. We do not know for example the number
$M$ of gaussians in (25).

As suggested by Guyon et al. (1992) and Vapnik (1992), the method to fix
the parameters in the expression (26) is to minimize the sum of a pair of
competing terms. The former is again the standard error given by eq. (22),
and decreases monotonically as the number of parameters is increased.
The latter measures the complexity of the model and increases by
increasing the number of parameters. Therefore there is
an optimal compromise which minimizes the sum.
It is possible to demonstrate (Amari et al. 1997)
that we obtain this compromise by using $N-k$ data
and the average error made on the patterns left out. This method is
called {\it leave-k-out cross validation}. When $N$ is small, the
most reasonable choice is $k=1$ ({\it leave-one-out}).

Since in our case the value of $N$ is small we apply the {\it leave-one-out
cross-validation method} in order to fix the parameters in (26). The average errors
made on the left data point are reported in Table 1 related to several values of
$M$.

\begin{table}[t]
\begin{center}
\begin{tabular}{ll}
\hline
$M$ & $Error$  \\
\hline
$1$  & $0.9131$   \\ 
\hline
$2$  & $0.4988$   \\ 
\hline
$3$  & $0.7545$   \\ 
\hline
$4$  & $0.8080$   \\ 
\hline
$5$  & $0.9097$   \\ 
\hline
\end{tabular}
\caption{Average errors
made on the left data points following the {\it leave-one-out cross validation} 
method for several values of $M$.}
\end{center}
\end{table}

Consequently, the best model occurs for $M=2$ with values of the parameters
appearing in eq. (26) given by eq. (3).

\end{document}